\documentclass[aps,pra,twocolumn,superscriptaddress,nofootinbib]{revtex4-2}
\usepackage{physics}
\usepackage[dvipdfmx]{graphicx}
\usepackage{here}
\usepackage{amsmath}
\usepackage{amssymb}
\usepackage{bm}
\usepackage{wrapfig}
\usepackage{ascmac}
\usepackage{mathrsfs}
\usepackage{color}
\usepackage{comment}
\usepackage[normalem]{ulem}
\usepackage{amsthm}
\usepackage{mathtools}

\theoremstyle{plain}

\theoremstyle{definition}

\theoremstyle{remark}

\theoremstyle{plain}
\newtheorem*{thm*}{Theorem}
\newtheorem*{lem*}{Lemma}
\newtheorem*{prop*}{Proposition}
\newtheorem*{cor*}{Corollary}
\newtheorem*{conj*}{Conjecture}

\theoremstyle{definition}
\newtheorem*{ass*}{Assumption}
\newtheorem*{dfn*}{Definition}

\theoremstyle{remark}
\newtheorem*{rem*}{Remark}

\newcommand{\im}{{\rm i}}

\usepackage{comment}
\usepackage{hyperref}
\usepackage[normalem]{ulem}
\hypersetup{colorlinks=true,citecolor=blue,linkcolor=blue,urlcolor=blue}

\begin{document} 
\title{Quantum Mpemba effect in long-range spin systems} 
\author{Shion Yamashika}
\affiliation{Department of Engineering Science, The University of Electro-Communications, Tokyo 182-8585, Japan.}
\author{Filiberto Ares}
\affiliation{SISSA and INFN, via Bonomea 265, 34136 Trieste, Italy.}

\begin{abstract}
One of the manifestations of the quantum Mpemba effect (QME) is that a tilted ferromagnet exhibits faster restoration of the spin-rotational symmetry after a quantum quench when starting from a larger tilt angle. This phenomenon has recently been observed experimentally in an ion trap that simulates a long-range spin chain. 
However, the underlying mechanism of the QME in the presence of long-range interactions remains unclear. Using the time-dependent spin-wave theory, we investigate the dynamical restoration of the spin-rotational symmetry and the QME in generic long-range spin systems. 
We show that quantum fluctuations of the magnetization drive the restoration of symmetry by melting the initial ferromagnetic order and are responsible for the QME. We find that this effect occurs across a wide parameter range in long-range systems, in contrast to its absence in some short-range counterparts. 
\end{abstract}

\maketitle

\textit{Introduction.}---
Non-equilibrium many-body
quantum systems exhibit a rich phenomenology absent in equilibrium, including quantum thermalization, anomalous transport, and emergent dynamical phases~\cite{Polkovnikov:2010yn,Gogolin:2015gts,Calabrese:2016psf,Heyl:2017blm}. Today we possess not only a deeper understanding of these phenomena but also significant experimental insights, thanks to remarkable progress in controllable quantum simulators that enable their realization in the laboratory~\cite{Blatt2012, Georgescu:2013oza, Gross2017, Hofstetter2018}. 
\par 
Among the most striking and counterintuitive non-equilibrium phenomena, the Quantum Mpemba Effect (QME) has recently garnered considerable attention. As in its classical counterpart~\cite{Mpemba_1969, Lu2017, Klich2019, Kumar2020, Bechhoefer2021, Kumar2022, VanVu2025}, it describes the situation in which a system initially farther from 
equilibrium relaxes faster than one initially closer to it. Several versions of this effect have been studied in recent years across a variety of quantum systems~\cite{Nava2019, Carollo2021, Ares:2022koq, Hayakawa2023, Moroder2024, Nava2024, Strachan2025, Longhi2025} and, beyond its theoretical interest, it is also attracting attention as a practical tool for quantum control and state preparation~\cite{westhoff2025fast,boubakour2025dynamical}; see also the reviews~\cite{Teza2025review, Ares2025review,ylz-25-rev}. 

A particularly relevant case is that of a quantum quench in a closed system~\cite{Ares:2022koq}. In this scenario, the system follows a unitary time evolution and the QME arises purely from quantum correlations. In closed quantum systems, the relaxation to equilibrium occurs locally. An ideal proxy for monitoring this process is symmetry: the system starts in an nonequilibrium state that breaks a global internal symmetry and undergoes a unitary evolution that respects such symmetry. In general, a subsystem eventually relaxes to a symmetric state. In this context, the QME manifests when the symmetry is locally restored faster for the initial state that breaks it more.
\par 
This form of the QME has been experimentally observed in a trapped-ion quantum simulator~\cite{Joshi:2024sup} --- see~\cite{Shapira2024, Zhang2025} for experimental realizations of other quantum versions of the effect. The ion-trap is initially prepared in different tilted ferromagnetic states, which break to a different extent the rotational spin symmetry around the $z$-axis. The system then evolves, simulating the dynamics of a long-range
interacting spin-$1/2$ Hamiltonian that respects the rotational symmetry. The QME was probed through direct measurements along the time evolution of the entanglement asymmetry --- a quantum information based observable that quantifies the degree of symmetry breaking in a subsystem. 

Despite the experimental success, a theoretical explanation for the occurrence of the QME in that setup is still lacking. Here, we aim to bridge this gap. The QME in closed quantum systems via symmetry restoration is actually only fully understood in integrable systems in terms of the quasiparticle picture: It occurs when, in the state that initially breaks more the symmetry, most of the charge is transported by the fastest entangled pairs of quasiparticles created in the quench~\cite{Rylands:2023yzx, Bertini2024, Murciano2024, Chalas2024, Yamashika2024, Klobas2024, Rylands:2024fio, yamashika-2025, Caceffo2024, Ares2025}. However, this picture cannot be generally applied~\cite{Banerjee2024, Liu2024-2, DiGiulio2025, Benini2025, Klobas2025, Yu2025-2, Gibbins2025} --- in particular, to the long-range spin model of the experiment in Ref.~\cite{Joshi:2024sup}. A unified framework that explains the QME in ergodic systems remains elusive. Some mechanisms have been proposed for $U(1)$-symmetric random unitary circuits~\cite{Liu2024, Turkeshi2024, Yu2025, Foligno2025} as well as in Hamiltonian systems devoid of any symmetry~\cite{Bhore2025, Ares2025-2}. 

In this Letter, employing a semiclassical approach based on the time-dependent spin-wave approximation, we investigate the microscopic mechanism of dynamical symmetry restoration and identify the origin of the QME in generic $U(1)$-symmetric long-range interacting spin systems after a quench from ferromagnetic ordered states. 
Understanding these systems is important not only because of their relevance to QME experiments, but also due to their fundamentally distinct dynamics compared to short-range systems, including prethermalization and anomalous entanglement propagation~\cite{Hauke2013prl,Schachenmayer2013prx,Richerme:2014qla, Jurcevic2014, Buyskikh-2016, Frerot2018, Cevolani2018, Joshi2022, Colmenarez:2020juf, Defenu2023, Defenu2024}.
\par 
\textit{Setting.}---
Let us consider the quench dynamics of a $d$-dimensional lattice of $N$ spin-$s$ that is initially prepared in the tilted ferromagnetic state,
\begin{align}
\ket{\mathrm{TF}(\theta,\phi)}
=
e^{-\im \phi \hat{M}^z}
e^{-\im \theta \hat{M}^y} \ket{\uparrow,\uparrow,...,\uparrow},
\label{eq:tilted-ferro} 
\end{align}
where $\hat{M}^\mu=\sum_{i=1}^N\hat{S}_i^\mu$ is the total magnetization operator along the $\mu$ axis and $\hat{\mathbf{S}}_i=(\hat{S}_i^x,\hat{S}_i^y,\hat{S}_i^z)$ is the spin-$s$ operator in $i$-th site. 
We denote as $\ket{\uparrow(\downarrow)}_i$ the eigenstates of $\hat{S}_i^z$ with eigenvalue $\pm s$, respectively. When $\theta=0$ or $\pi$, the state~\eqref{eq:tilted-ferro} is invariant under global spin rotations around the $z$-axis, whereas it breaks this symmetry for $0<\theta<\pi$.  

The system evolves as $\ket{\Psi_t}=e^{-\im t H}\ket{\mathrm{TF}(\theta,\phi)}$, with a Hamiltonian $H$ invariant under global spin rotations around the $z$-axis. If we further assume that $H$ is symmetric under translations in the lattice, then it can be written, without loss of generality, as 
\begin{multline}
H=-\sum_{i,j=1}^N J(|\mathbf{r}_i-\mathbf{r}_j|)[\hat{s}_i^x \hat{s}_j^x +\hat{s}_i^y \hat{s}_j^y
\\
+(1-\Delta)\hat{s}_i^z \hat{s}_j^z]-h\sum_{i=1}^N \hat{s}_i^z,
    \label{eq:XXZ}
\end{multline}
where $\hat{\mathbf{s}}_i=\hat{\mathbf{S}}_i^\mu/s$ are the normalized spin-$s$ operators, $\mathbf{r}_i$ is the $d$-dimensional vector identifying the position of the $i$-th spin, $J(|\mathbf{r}_i-\mathbf{r}_j|)$ is the strength of interactions between the $i$-th and the $j$-th spins, $\Delta$ is the anisotropy parameter of the interactions, and $h$ is an external field in the $z$ direction. 
In the experiment in Ref.~\cite{Joshi:2024sup}, the ion trap simulates a $d=1$ chain of $N=12$ spins $s=1/2$ prepared in states of the form~\eqref{eq:tilted-ferro} with $\phi=0$  and different $\theta$ that evolves with the Hamiltonian~\eqref{eq:XXZ}, with $\Delta=1$, $h=0$, $J(|\mathbf{r}|)\propto |\mathbf{r}|^{-\alpha}$, and $\alpha\approx 1$.
\par
As in the experiment, we are interested in the fate of the rotational symmetry broken by the initial state \eqref{eq:tilted-ferro} after the quench. To this end, we consider a subsystem $A$ of $N_A$ contiguous spins. In this subsystem, the spin rotations around the $z$-axis are generated by $\hat{M}_A^z = \sum_{i\in A} \hat{S}_i^z$. In the limit $N\to\infty$, the reduced density matrix of subsystem $A$, $\rho_A(t) = \Tr_{\bar{A}}[\ket{\Psi_t}\!\bra{\Psi_t}]$, relaxes to a statistical ensemble determined by the Hamiltonian~\eqref{eq:XXZ} that respects the rotational symmetry initially broken, i.e., $[\rho_A(t\to\infty),\hat{M}_A^z]=0$ \cite{Rigol:2007juv,Deutsch:1991msp,Srednicki:1994mfb}. 
When $\rho_A$ commutes with $\hat{M}_A^z$, it is block diagonal in the eigenbasis of $\hat{M}_A^z$. 
Based on this property, we can quantify the extent the symmetry is broken in $A$ and monitor its restoration with the entanglement asymmetry \cite{Ares:2022koq}, defined as
\begin{align}
\Delta S_A = \ln(\Tr_A[\rho_A^2]) - \ln(\Tr_A[\tilde{\rho}_A^2]), \label{eq:EA}
\end{align}
where $\tilde{\rho}_A = \sum_m \Pi_m \rho_A \Pi_m$ denotes the symmetrized reduced density matrix, and $\Pi_m$ is the projection operator onto the eigenspace of $\hat{M}_A^z$ with eigenvalue $m$. 
The entanglement asymmetry \eqref{eq:EA}  is positive semidefinite, $\Delta S_A \geq 0$, and vanishes if and only if $\rho_A$ is symmetric, that is, $\Delta S_A=0$ iff $[\rho_A, \hat{M}_A^z] = 0$ \cite{Ares:2022koq}. 
In Ref.~\cite{Joshi:2024sup}, the entanglement asymmetry~\eqref{eq:EA} was directly measured in the ion-trap quantum simulator by applying the randomized-measurement toolbox \cite{Elben:2022jvo,Huang:2020tih}.
\par 
The conditions for the occurrence of the QME can be formulated in terms of $\Delta S_A$. 
Let $\rho_{A,i}(t)$ ($i=1,2$) be the reduced density matrices of $A$ for two different initial states, both evolved under the same Hamiltonian, with $\Delta S_{A,i}(t)$ their entanglement asymmetry at time $t$. Then, the QME is said to occur if (i) $\rho_{A,1}(0)$ initially breaks more the symmetry than $\rho_{A,2}(0)$, $\Delta S_{A,1}(0)>\Delta S_{A,2}(0)$; and (ii) after a certain time $t_M$ (Mpemba time), $\rho_{A,1}(t)$ is more symmetric than $\rho_{A,2}(t)$, i.e., $\Delta S_{A,1}(t)<\Delta S_{A,2}(t)$ $\forall t>t_M$.
\begin{figure*}
\includegraphics[width=0.85\textwidth]{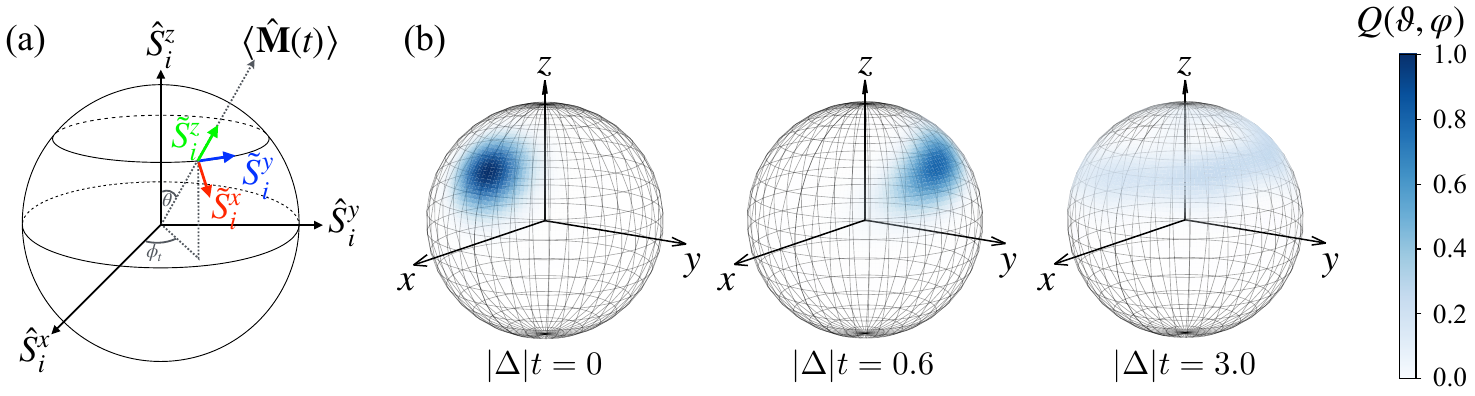}
\caption{(a) Relation between spin operators ($\hat{S}_i^x,\hat{S}_i^y,\hat{S}_i^z$) in the laboratory frame and those ($\tilde{S}_i^x,\tilde{S}_i^y,\tilde{S}_i^z$) in the rotated frame. In the rotated frame, $\tilde{S}_i^x$ and $\tilde{S}_i^y$ correspond to the spin fluctuation components in the polar and azimuthal directions, while $\tilde{S}_i^z$ is the component in the direction of the precessing magnetization $\langle \hat{\mathbf{M}}(t)\rangle$.
(b) Quasidistribution function $Q(\vartheta,\varphi)=|\bra{\Psi_t}\ket{\mathrm{TF}(\vartheta,\varphi)}|^2$ in spherical coordinates for the quench dynamics of a long-range spin-1/2 chain. We set $N=24$, $\Delta=1$, $\alpha=0$, and initial tilted angle $\theta=0.3\pi$.}
\label{fig:RotatedFrame}
\end{figure*}
\par 
\textit{Entanglement asymmetry from time-dependent spin-wave theory.}---
We can determine when and how conditions (i) and (ii) are met within our setup by calculating the time evolution of the entanglement asymmetry.
To this end, we employ the time-dependent spin-wave theory~\cite{Ruckriegel:2012,Lerose-2018,Lerose-2019,Lerose-2020}, an effective theory that treats fluctuations of spins in classically ordered states as perturbations.
In our setup, the magnetization $\langle \hat{\mathbf{M}}(t)\rangle =\bra{\Psi_t} (\hat{M}^x,\hat{M}^y,\hat{M}^z)^T \ket{\Psi_t}$ precesses around the $z$-axis since its $z$-component is conserved by the Hamiltonian~\eqref{eq:XXZ}~\cite{SM}. \nocite{Jutho-2016, Jutho-Cirac-2011,Pirvu_2010} 
To apply the time-dependent spin-wave theory, we move to the rotated frame in which $\langle \hat{\mathbf{M}}(t)\rangle$ is aligned with the $z$-axis. The spin operators in that frame, $\tilde{S}_i^\mu$, are related to the original ones by $\tilde{S}_i^\mu=R_t \hat{S}_i^\mu R_t^\dag$,  where $R_t = e^{-\im \phi_t \hat{M}^z} e^{-\im \theta\hat{M}^y}$ and $\phi_t$ is the azimuthal angle of the magnetization at time $t$, see Fig.~\ref{fig:RotatedFrame}~(a) and~\cite{SM}. 
We then perform a semiclassical expansion of the spin operators in the rotated frame via the Holstein-Primakoff transformation~\cite{Holstein:1940zp}, 
\begin{align}
    \tilde{S}_i^x \simeq \sqrt{\frac{s}{2}}(b_i+b_i^\dag), 
    \tilde{S}_i^y \simeq \im \sqrt{\frac{s}{2}} (b_i^\dag-b_i),
    \tilde{S}_i^z=s-b_i^\dag b_i,
        \label{eq:HP}
\end{align}
where $b_i$ ($b_i^\dag$) is a bosonic annihilation (creation) operator associated with the $i$-th spin. 
In the bosonic picture, the initial tilted-ferromagnetic state corresponds to the bosonic vacuum state, and the quantum fluctuations of spins are described by bosonic excitations, which are referred to as spin waves.
We approximate the Hamiltonian~\eqref{eq:XXZ} in the rotated frame, $ \tilde{H}=H+\im R_t \partial_t R_t^\dag$ \cite{Ruckriegel:2012,Lerose-2018,Lerose-2019,Lerose-2020}, by a quadratic expansion in $\boldsymbol{b}_i=(b_i^\dag ,b_i)$ and obtain
\begin{align}
    \Tilde{H}\simeq 
    \frac{1}{2s}
    \sum_\mathbf{k} (\xi_\mathbf{k}b_\mathbf{k}^\dag b_\mathbf{k} +\kappa_\mathbf{k} b_\mathbf{k}^\dag b_\mathbf{-k}^\dag+\mathrm{H.c.})
    \label{eq:H_eff},
\end{align}
where $b_\mathbf{k}=\sum_{i}e^{-\im \mathbf{k}\cdot \mathbf{r}_i}b_i/\sqrt{N}$ is the annihilation operator for the spin waves with momentum $\mathbf{k}$ and 
\begin{gather}
\xi_\mathbf{k}=2(\tilde{J}_\mathbf{0}-\tilde{J}_\mathbf{k})+\tilde{J}_\mathbf{k} \Delta \sin^2 \theta,\\
\kappa_\mathbf{k}=\tilde{J}_\mathbf{k}\Delta \sin^2 \theta,~
\tilde{J}_\mathbf{k}=\sum_{i=1}^NJ(|\mathbf{r}_i|)e^{\im \mathbf{k}\cdot \mathbf{r}_i}.
\end{gather}
The quadratic expansion performed in deriving Eq.~\eqref{eq:H_eff} is valid when the spin waves are dilute, namely, when the spin fluctuations are small. 
As explicitly shown in~\cite{SM}, the mode occupation number of spin waves $n_{\mathbf{k}}$ evolves in time as 
 \begin{align}
n_\mathbf{k}(t)= \frac{\Delta^2 \tilde{J}_\mathbf{k}^2 \sin^4\theta}{\omega_\mathbf{k}^2} \sin^2 \qty(\frac{\omega_\mathbf{k}t}{s}),  
\label{eq:n_k}
 \end{align}
where $\omega_\mathbf{k}=\sqrt{\xi_\mathbf{k}^2 -\kappa_\mathbf{k}^2}$ is the Bogoliubov dispersion relation for the effective Hamiltonian \eqref{eq:H_eff}.
Provided $\Im[\omega_\mathbf{k}]=0~\forall \mathbf{k}$, the right-hand side of Eq.~\eqref{eq:n_k} can be bounded above by $O[(t/s)^2]$ and thus the spin wave theory is valid up to the Ehrenfest time $t_\mathrm{Ehr}=O(\sqrt{sN})$, which diverges in either the thermodynamic limit ($N\to\infty$) or in the classical spin limit ($s\to\infty$), or both. 
When $\Im[\omega_\mathbf{k}]\neq0$ for some $\mathbf{k}$, the corresponding mode occupation numbers grow exponentially, indicating that the precessing ferromagnetic order is dynamically unstable and hence the spin-wave theory is not applicable. 
We thus assume $\Im[\omega_\mathbf{k}]=0~\forall \mathbf{k}$ for the theory to be valid. 
This condition is satisfied when the interactions between spins are sufficiently long ranged, so that the precessing ferromagnetic order remains stable during the time evolution, see \cite{SM} for a more quantitative discussion.
\begin{figure*}
\raggedright
\includegraphics[width=\textwidth]{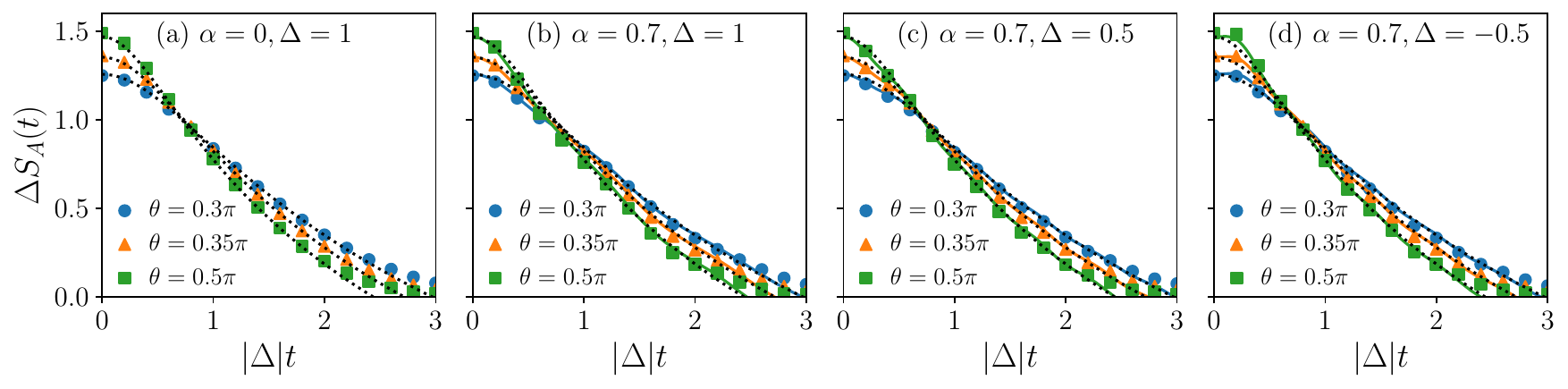}
\caption{
Time evolution of the entanglement asymmetry after a quench from different tilted ferromagnetic states in the long-range spin-$1/2$ chain~\eqref{eq:XXZ} with $J(|\mathbf{r}|)\propto|\mathbf{r}|^{-\alpha}$ and $h=0$. 
The parameters $(\alpha, \Delta)$ are set to $(0, 1)$, $(0.7, 1)$, $(0.7, 0.5)$, and $(0.7, -0.5)$ in panels (a), (b),
(c), and (d), respectively.
The symbols are the exact asymmetry obtained with ED. The solid curves are the prediction of the time-dependent spin-wave theory (Eq.~\eqref{eq:EA_analytic}) and the dotted curves (Eq.~\eqref{eq:EA_long-range-limit}) are the dominant contribution of the zero momentum spin-waves. 
The crossings between the curves for the different values of $\theta$ indicate the occurrence of the QME.  
We set $N=24$ and $N_A=6$ in all the plots.
}
\label{fig:EA}
\end{figure*}
\par 
As long as the spin-wave theory is valid, the dynamics of the system is described by the quadratic Hamiltonian~\eqref{eq:H_eff}. 
This implies that the time-evolved reduced density matrix $\rho_A(t)$ satisfies Wick's theorem and, therefore, it is univocally characterized by the mean vector $\langle \boldsymbol{b}_i\rangle$ and the covariance matrix $\Gamma$, which is a $2N_A$-dimensional block matrix with entries
\begin{align}
    \Gamma_{ij}\!=\!
    \mqty[\langle \{b_i^\dag\!-\!\langle b_i^\dag\rangle,b_j^\dag\!-\!\langle b_j^\dag \rangle\}\rangle&\langle \{b_i^\dag\!-\!\langle b_i^\dag\rangle,b_j\!-\!\langle b_j \rangle\}\rangle
    \\
    \langle \{b_i\!-\!\langle b_i\rangle,b_j^\dag\!-\!\langle b_j^\dag \rangle\}
    \rangle & 
    \langle \{b_i\!-\!\langle b_i\rangle,b_j\!-\!\langle b_j \rangle\}\rangle
    ],
\end{align}
with $i,j\in A$ and $\{X,Y\}=XY+YX$.
This property allows us to calculate the entanglement asymmetry~\eqref{eq:EA} by standard bosonic Gaussian techniques~\cite{Peschel:2002yqj,Botero:2004vpl,Banchi:2015rmr,yamashika-2025}. 
In particular, we find that, for large subsystems $N_A\gg1$, it is given by~\cite{SM}
\begin{align}
\Delta S_A(t)\simeq \frac{1}{2}\ln(s\pi \sin^2 \theta[\boldsymbol{u}\cdot \Gamma(t)^{-1}\boldsymbol{u}]), 
\label{eq:EA_analytic}
\end{align}
where $\boldsymbol{u}$ is the $2N_A$-dimensional vector whose elements are $u_i=(-1)^{i}$.

Equation~\eqref{eq:EA_analytic} can be expanded in terms of the density of spin waves $n_\mathbf{k}$, see~\cite{SM}, as
\begin{align}
\Delta S_A (t)\simeq \frac{1}{2} \ln(\frac{2s\pi N_A\sin^2 \theta}{1+4 n_\mathbf{0}(t) f_A(f_A-1)})+O\qty(\frac{n_\mathbf{k\neq0}(t)}{n_\mathbf{0}(t)}), 
\label{eq:EA_long-range-limit}
\end{align}
where $f_A=N_A/N$ is the fraction of spins inside subsystem $A$. At $t=0$, since $n_{\mathbf{k}}(t)=0$, we have
\begin{align}
    \Delta S_A(t=0)\simeq 
    \frac{1}{2}\ln(s\pi N_A \sin^2\theta). 
    \label{eq:EA_short time}
\end{align}
This is the entanglement asymmetry of the tilted-ferromagnetic state~\eqref{eq:tilted-ferro} for large subsystems $N_A\gg1$, which was the derived for spin-1/2 chains in Ref.~\cite{Ares:2022koq}. 
Observe that the degree the symmetry is broken at $t=0$ increases monotonically as the polar angle $\theta$ approaches $\pi/2$, at which it is maximal.

For $t>0$, the first term in 
the expansion in Eq.~\eqref{eq:EA_long-range-limit} only depends on the number of spin waves with zero momentum and it generally provides the dominant contribution to the entanglement asymmetry. 
The reason is that, given Eq.~\eqref{eq:n_k}, the ratio between the number of spin waves with zero momentum and those with finite momenta is bounded above as
\begin{align}
\frac{n_\mathbf{k\neq 0}(t)}{n_\mathbf{0}(t)}=\qty(\frac{\tilde{J}_\mathbf{k}}{\tilde{J}_\mathbf{0}})^2 \frac{\sin^2(\omega_\mathbf{k}t/s)}{(\omega_\mathbf{k}t/s)^2} \leq
O[(s/t)^2].
\label{eq:n_k/n_0}
\end{align} 
Therefore, $n_{\mathbf{k\neq0}}(t)$ is comparable to $n_{\mathbf{0}}(t)$ only at short times, and the subsequent evolution of the entanglement asymmetry is mainly governed by the zero-momentum spin waves. 
In particular, in the long-range interaction limit $J(|\mathbf{r}|) = \mathrm{const.}$, Eq.~\eqref{eq:EA_long-range-limit} is accurate even at short times, as the emergent permutation symmetry of spins in this limit forbids the excitation of spin waves with finite momenta during the time evolution. A more quantitative discussion on the contribution of finite-momentum spin waves is provided in~\cite{SM}. 

According to Eqs.~\eqref{eq:EA_long-range-limit} and~\eqref{eq:n_k/n_0}, in the classical limit $s\gg 1$, the entanglement asymmetry is constant after the quench, suggesting that the quantum fluctuations of the spin are the crucial ingredient for the restoration of the symmetry, as we will see later.

We check the results above in Fig.~\ref{fig:EA}, where we plot the time evolution of the entanglement asymmetry taking the Hamiltonian in Eq.~\eqref{eq:XXZ} with $J(|\mathbf{r}|)=1/(K|\mathbf{r}|^{\alpha})$, where $K=N^{-1}\sum_{j\neq j'}|\mathbf{r}_{j}-\mathbf{r}_{j'}|^{-\alpha}$ is the Kac rescaling factor, $h=0$, $d=1$, and $s=1/2$. We consider different exponents $0\leq \alpha\leq 1$, for which the long-range interaction decays slowly enough so that the condition $\Im[\omega_\mathbf{k}]=0~\forall\mathbf{k}$ is satisfied and the spin-wave theory is valid, see \cite{SM}. 
We numerically compute the entanglement asymmetry by exact diagonalization (ED), comparing it with the predictions of Eqs.~\eqref{eq:EA_analytic} (solid curves) and~\eqref{eq:EA_long-range-limit} (dotted curves). 
The entanglement asymmetries decay to zero and intersect only once for any pair of initial states~\eqref{eq:tilted-ferro} and $\Delta\neq 0$, indicating the occurrence of the QME. As expected, Eq.~\eqref{eq:EA_long-range-limit} deviates from~\eqref{eq:EA_analytic} at short times for $\alpha>0$ due to the contribution of the finite momentum spin waves, which are suppressed in the infinite-range case ($\alpha=0$, panel (a)).

\textit{Generality of the QME.}---
Equipped with the previous results, we can show that the QME generally occurs in $U(1)$-symmetric long-range spin systems in a quench from ferromagnetic ordered states. 

Let us consider two systems prepared in the tilted-ferromagnetic states with polar angles $\theta_i\in(0, \pi/2]$ ($i=1,2$). 
On one hand, according to Eq.~\eqref{eq:EA_short time}, condition (i) is satisfied when $\theta_1>\theta_2$.
On the other hand, condition (ii) is equivalent to the existence of a Mpemba time $t_M > 0$ at which $\Delta S_{A,1}(t_M) = \Delta S_{A,2}(t_M)$, since Eqs.~\eqref{eq:n_k} and~\eqref{eq:EA_long-range-limit} imply that $\Delta S_{A,1}(t)$ and $\Delta S_{A,2}(t)$ intersect only once, as also evidenced numerically in Fig.~\ref{fig:EA}.
Solving $\Delta S_{A,1}(t_M)=\Delta S_{A,2}(t_M)$ using Eqs.~\eqref{eq:n_k} and~\eqref{eq:EA_long-range-limit}, we obtain 
\begin{align}
    t_M
    =
    \frac{s}{2\sqrt{f_A(1-f_A)}|\tilde{J}_\mathbf{0}\Delta \sin\theta_1\sin\theta_2|}.
\end{align}
When condition (i) is met, i.e., $\theta_1>\theta_2$, the Mpemba time $t_M$ given by this equation is typically finite and smaller than the Ehrenfest time $t_\mathrm{Ehr}=O(\sqrt{sN})$, beyond which the time-dependent spin-wave theory becomes unreliable. 
Exceptions occur in exotic cases, such as when $\tilde{J}_\mathbf{0}=0$, in which the interactions between spins can be ferromagnetic or antiferromagnetic depending on the distance, or the isotropic point $\Delta=0$, at which the spin waves are not excited as shown by Eq.~\eqref{eq:n_k}.
We thus conclude that conditions (i) and (ii) are satisified and thereby the QME occurs in the quench dynamics of a broad variety of symmetric long-range spin systems starting from tilted-ferromagnetic states. In the same quenches, but
taking nearest-neighbor interactions in~\eqref{eq:XXZ}, the QME only occurs in the regime $0<\Delta<2$~\cite{Ares:2022koq, Rylands:2024fio}.
\par 
\textit{Physical interpretation.}---
The analytic prediction~\eqref{eq:EA_long-range-limit} also provides a clear interpretation of the mechanism underlying the dynamical restoration of spin rotations in the presence of long-range interactions. According to this expression, the entanglement asymmetry decreases as the number of zero-momentum spin waves, $n_{\mathbf{0}}$, increases. 
By performing the inverse of the Holstein-Primakoff transformation~\eqref{eq:HP}, $n_\mathbf{0}$ can be expressed in terms of the spin operators in the rotated frame,
\begin{align}
n_\mathbf{0} = \frac{\mathrm{Var}\qty(\sum_{i=1}^N\tilde{S}_i^x)+\mathrm{Var}\left( \sum_{i=1}^N \tilde{S}_i^y \right)}{2sN} - \frac{1}{2},
\label{eq:n_k=0 Var}
\end{align}
where \( \mathrm{Var}(\cdot) = \langle \cdot^2 \rangle - \langle \cdot \rangle^2 \) denotes the variance. 
As illustrated in Fig.~\ref{fig:RotatedFrame}~(a), $\tilde{S}_i^x$ and $\tilde{S}_i^y$ are the polar and azimuthal components of the precessing magnetization $\langle \hat{\mathbf{M}}(t)\rangle$, respectively.
Therefore, Eq.~\eqref{eq:n_k=0 Var} indicates that the excitation of zero-momentum spin waves after the quench corresponds to the generation of fluctuations in the transverse components of the magnetization. More specifically, since 
\begin{align}
    \frac{\mathrm{Var}\qty(\sum_{i=1}^N\tilde{S}_i^x)}{sN}&=\frac{1}{2},\\
    \frac{\mathrm{Var}\qty(\sum_{i=1}^N\tilde{S}_i^y)}{sN}&=\frac{1}{2}+\tilde{J}_\mathbf{0}^2\Delta^2 \sin^4\theta (t/s)^2,
    \label{eq:Var(S^y)}
\end{align}
the magnetization fluctuations only increase in the azimuthal direction during the time evolution. 

At $t=0$, the magnetization $\langle \hat{\mathbf{M}}(t)\rangle$ points in a direction away the $z$
axis and the symmetry under rotations around this axis is broken. After the quench, the symmetric dynamics not only makes the magnetization to precess around the axis but it also increases the quantum fluctuations in the azimuthal direction. As visualized in Fig.~\ref{fig:RotatedFrame}~(b) in terms of the quasidistribution function $Q(\vartheta,\varphi)=|\bra{\Psi_t}\ket{\mathrm{TF}(\vartheta,\varphi)}|^2$, these fluctuations cause the direction of the precessing magnetization along its classical trajectory to become increasingly uncertain, ultimately leading to the restoration of the rotational symmetry around the $z$-axis in the long time limit. 
\par 
From this perspective, the QME in symmetric long-range spin systems can be regarded as a property of ferromagnetic states: the closer the polar angle $\theta$ is to $\pi/2$, i.e. the more the symmetry is initially broken, the faster the quantum uncertainty of the azimuthal magnetization grows after the quench, causing the symmetry to be restored sooner. 
This mechanism is distinct from that in the short-range integrable counterparts of Hamiltonian~\eqref{eq:XXZ}, where 
the emergence of the QME is attributed to charge transport properties of the system~\cite{Rylands:2023yzx}.   
\par 
\textit{Conclusions}.---
We investigated the dynamical restoration of rotational symmetry and the emergence of the QME in generic long-range spin systems, following quenches from ferromagnetic ordered states. To monitor the extent to which the symmetry is broken in a subsystem, we used the entanglement asymmetry. By applying the time-dependent spin-wave approximation, we derived analytical expressions for this observable and demonstrated that the QME generally occurs in such quenches. 

We found that the time evolution of the entanglement asymmetry is governed by the zero-momentum spin waves generated after the quench. These excitations are responsible for increasing the quantum fluctuations of
the global magnetization, leading to the melting of the initial ferromagnetic order and the eventual restoration of rotational symmetry. The rate at which the zero-momentum spin waves are generated increases with the asymmetry of the initial states, thereby explaining the emergence of the QME.
These results furnish a theoretical background for the recent experimental observations of the QME~\cite{Joshi:2024sup}.

Our approach can be readily applied to study other symmetries—for example, spin parity in long-range Ising chains. It may also be extended to examine the QME in dissipative and monitored long-range spin systems~\cite{Seetharam2022, Delmonte2025}. However, it is limited to classically ordered initial states. In the case of disordered initial states, Gaussian approximations of the system dynamics can still be obtained, as we did here, albeit using different formalisms~\cite{Senese2024}. The exploration of these directions lies beyond the scope of the present work and will be addressed in future studies.
\par 
\ 
\par 
\textit{Note added.---} After this manuscript was submitted, Ref.~\cite{xu2025exp} reported an independent experimental observation of the quantum Mpemba effect in a superconducting processor implementing $U(1)$ symmetric long-range spin chains, consistent with the mechanism proposed here.
\par 
\ 
\par
\textit{Acknowledgments.---} 
SY thanks D. Kagamihara for fruitful discussions. 
FA acknowledges support from European Union-NextGenerationEU, in the framework of the PRIN 2022 Project HIGHEST no. 2022SJCKAH\_002 and from ERC under Advanced Grant No. 101199196 (MOSE). SY acknowledges support from Grant-in-Aid
for Young Scientists (Start-up) No. 25K23355, University
of Electro-Communications, and the computational resources provided by S. Tsuchiya.

\bibliographystyle{apsrev4-2}
\bibliography{refs}  

\onecolumngrid
\newpage 
\newcounter{equationSM}
\newcounter{figureSM}
\newcounter{tableSM}
\stepcounter{equationSM}
\setcounter{equation}{0}
\setcounter{figure}{0}
\setcounter{table}{0}
\setcounter{section}{0}
\makeatletter
\renewcommand{\theequation}{\textsc{sm}-\arabic{equation}}
\renewcommand{\thefigure}{\textsc{sm}-\arabic{figure}}
\renewcommand{\thetable}{\textsc{sm}-\arabic{table}}

\begin{center}
  {\large{\bf Supplemental Material for\\ ``Quantum Mpemba effect in long-range spin systems''}}
\end{center} 

In this supplemental material, we report some useful information complementing the main text:
\begin{itemize}
    \item In Sec.~\ref{SMsec:swt}, we discuss the time-dependent spin wave theory employed to obtain the analytic results presented in the main text. 
    \item In Sec.~\ref{sec:validity}, we discuss the validity of the spin-wave theory in the one-dimensional long-range spin-1/2 XXZ Hamiltonian with power-law decaying interactions. 
    \item In Sec.~\ref{SMsec:asymmetry}, we give the details of the derivation of Eq.~(10) of the main text. 
    \item In Sec.~\ref{SMeq:expansion_asymm}, we describe how to derive Eq.~(11) of the main text and determine its first order correction.
    \item In Sec.~\ref{SMsec:numerics}, we present numerical results for large system sizes that are beyond the reach of brute-force ED.
\end{itemize}

\section{Time-dependent spin-wave theory}\label{SMsec:swt}

Here, we describe the formalism of the time-dependent spin wave theory and derive the effective Hamiltonian~(5) of the main text. 
We consider a $d$-dimensional $N$ spin-$s$ system that is invariant under global spin rotations around the $z$-axis and spatial translations in the spin lattice. 
The corresponding Hamiltonian is 
\begin{align}
    H=-\sum_{i,j=1}^N
    J(\abs{\mathbf{r}_i-\mathbf{r}_j})[\hat{s}_i^x\hat{s}_j^x+\hat{s}_i^y\hat{s}_j^y+(1-\Delta)\hat{s}_i^z \hat{s}_j^z]-h\sum_{i=1}^N \hat{s}_i^z, 
    \label{eq:XXZ_sp}
\end{align}
where $\hat{s}_i^\mu=\hat{S}_i^\mu/s$ are the normalized spin-$s$ operators in the $\mu$-axis, $\mathbf{r}_i$ is the $d$-dimensional vector identifying the position of $i$-th spin, $J(\abs{\mathbf{r}_i-\mathbf{r}_j})$ is the strength of interactions between $i$-th and $j$-th spins, $\Delta$ is the anisotropy parameter of the interactions, and $h$ is the external field in $z$ direction. 
\par 
We move to the rotated frame in which the expectation value of the magnetization operator, $\langle \hat{\mathbf{M}}(t)\rangle$ is always aligned with the $z$-axis. 
If we denote as $\theta_t$ and $\phi_t$ the polar and azimuthal angles of the magnetization at time $t$, the spin operators in the rotated frame are 
\begin{align}
    \hat{S}_i^\mu\to\Tilde{S}_i=R_t \hat{S}_i^\mu R_t^\dag, 
\end{align}
where 
\begin{align}
    R_t=e^{-\im \phi_t \hat{M}^z} e^{-\im \theta_t \hat{M}^y}.
\end{align}
Accordingly, the Hamiltonian \eqref{eq:XXZ_sp} transforms into 
\begin{align}
    H\to \Tilde{H}=H+\im R_t\partial_t R_t^\dag. 
    \label{eq:H_tilde}
\end{align}
Now we perform a semiclassical expansion of the spin operators $\Tilde{S}_i^\mu$ in the Hamiltonian $\Tilde{H}$ by applying the Holstein-Primakoff transformation,
\begin{align}
    \Tilde{S}_i^x \simeq \sqrt{\frac{s}{2}}(b_i^\dag+b_i),~
    \Tilde{S}_i^y \simeq \im\sqrt{\frac{s}{2}}(b_i^\dag -b_i),~
    \Tilde{S}_i^z = s-b_i^\dag b_i. 
    \label{SMeq:HP}
\end{align}
Inserting Eq.~\eqref{SMeq:HP} into Eq.~\eqref{eq:H_tilde} and neglecting the cubic and higher order terms in $\boldsymbol{b}_i=(b_i,b_i^\dag)^T$, we obtain 
\begin{align}
    \Tilde{H}\simeq H_1+H_2+\mathrm{const.,}
    \label{eq:H_expand}
\end{align}
where 
\begin{align}
    \sqrt{s}H_1
    &=\frac{1}{\sqrt{2}}\sum_{i=1}^N
    (h\sin \theta_t -2\Delta \Tilde{J}_\mathbf{0}\sin \theta_t \cos \theta_t +s \sin\theta_t \partial_t \phi_t + \im \partial_t \theta_t )b_i^\dag +\mathrm{H.c.},
\end{align}
and
\begin{align}
    sH_2
    &=-\sum_{i,j=1}^N\Tilde{J}(\abs{\mathbf{r}_i-\mathbf{r}_j})
    \qty[
    \qty(1-\frac{\Delta \sin^2 \theta_t}{2})b_i^\dag b_j
    -
    \frac{\Delta \sin^2 \theta}{2}b_i^\dag b_j^\dag +\mathrm{H.c.}
    ]
    \nonumber\\
    &\quad \quad 
    +\sum_{i=1}^N 
    (h+s\partial_t \phi_t -2\Delta \Tilde{J}_0\cos \theta_t)
    \cos \theta_t b_i^\dag b_i + 2\Tilde{J}_\mathbf{0}\sum_i b_i^\dag b_i. 
\end{align}
In these expressions, we introduced $\Tilde{J}_\mathbf{k}=\sum_\mathbf{r}J(\abs{\mathbf{r}})e^{\im \mathbf{k\cdot r}}$.
\par 
Since the magnetization $\langle \hat{\mathbf{M}}(t)\rangle$ is always aligned with the $z$-axis in the rotated frame, its $x$ and $y$ components in this frame have to be zero during the time evolution, i.e, $\langle \Tilde{M}^x (t)\rangle = \langle \Tilde{M}^y(t)\rangle=0$. 
This constraint can be satisfied by choosing $\theta_t$ and $\phi_t$ such that the linear term in $\boldsymbol{b}_i$ in the Hamiltonian \eqref{eq:H_expand} vanishes, i.e., $H_1=0$.  
From this condition, we obtain the equations that determine the classical trajectory of the magnetization, 
\begin{align}
    s\partial_t \theta_t=0,\quad 
    s\partial_t \phi_t=2\Delta \tilde{J}_\mathbf{0}\cos \theta-h.
\end{align}
Solving these equations with the initial conditions $(\theta_0,\phi_0)=(\theta,\phi)$, which are the polar and azimuthal angles of the initial tilted ferromagnet, cf. Eq.~(1) in the main text, we obtain 
\begin{align}
    \theta_t=\theta,~\phi_t=(2\Delta \tilde{J}_\mathbf{0}\cos \theta-h)(t/s)+\phi. 
    \label{eq:theta and phi}
\end{align}
The expressions above clearly show that the magnetization undergoes precession following the quench. 
This originates from the fact that the Hamiltonian~(2) preserves the $z$ component of the magnetization, i.e. $[H,\hat{M}^z]=0$. 
\par 
Plugging Eq.~\eqref{eq:theta and phi} into Eq.~\eqref{eq:H_expand} and performing the Fourier transform, $b_\mathbf{k}=\sum_i e^{-\im \mathbf{k\cdot r_i}}b_i/\sqrt{N}$, we obtain the effective Hamiltonian in Eq.~(5) of the main text, 
\begin{align}
    \tilde{H}
    =
    \frac{1}{2s}\sum_\mathbf{k}\qty{\xi_\mathbf{k}b_\mathbf{k}^\dag b_\mathbf{k}+\kappa_\mathbf{k}b_\mathbf{k}^\dag b_\mathbf{k}+\mathrm{H.c.}},
    \label{eq:H_eff_sm}
\end{align}
with the definitions 
\begin{align}
    \xi_\mathbf{k}=2(\tilde{J}_\mathbf{0}-\tilde{J}_\mathbf{k})+\tilde{J}_\mathbf{k}\sin^2 \theta,~
    \kappa_\mathbf{k}=\tilde{J}_\mathbf{k}\Delta \sin^2 \theta.
\end{align}
\par 
Within the time-dependent spin-wave theory, the dynamics of the system is described by the effective Hamiltonian~\eqref{eq:H_eff_sm}, which is quadratic in terms of the bosonic operators $\boldsymbol{b}_i=(b_i,b_i^\dag)$. 
Therefore, the reduced density matrix $\rho_A$ satisfies Wick's theorem and can be univocally determined by the mean vector $\langle \boldsymbol{b}_i\rangle$ and the two-point correlation matrix $\Gamma_{ij}=\langle \{\boldsymbol{b}_i-\langle \boldsymbol{b}_i\rangle,\boldsymbol{b}_j^\dag-\langle \boldsymbol{b}_j^\dag\rangle \}\rangle$, where $i,j\in A$. 
We take this advantage to calculate the entanglement asymmetry. 
To this end, it is convenient to introduce the canonical operators 
\begin{align}
    q_i=\frac{b_i+b_i^\dag}{\sqrt{2}},~p_i=\frac{b_i-b_i^\dag}{\im \sqrt{2}}, 
\end{align}
which are Hermitian and satisfy the canonical commutation relations $[q_i,q_j]=[p_i,p_j]=0$ and $[q_i,p_j]=\im \delta_{i,j}$.
We define in terms of them the mean vector $\boldsymbol{s}$ and the covariance matrix $\Sigma$, whose entries are given by
\begin{align}
    \boldsymbol{s}_i=
    \left\langle \mqty( q_i  \\ p_i )\right\rangle,\,
    \Sigma_{ij}
    =
    \left\langle 
    \mqty(
    \{q_i-\langle q_i\rangle,q_j-\langle q_j\rangle \} 
    &
    \{q_i-\langle q_i\rangle,p_j-\langle p_j\rangle \} 
    \\
    \{p_i-\langle p_i\rangle,q_j-\langle q_j\rangle \} 
    &
    \{p_i-\langle p_i\rangle,p_j-\langle p_j\rangle \} 
    )
    \right\rangle.
\end{align}
The covariance matrix $\Sigma$ is related to the two-point correlation matrix $\Gamma$ through the unitary transformation
\begin{align}
    \Gamma=U\Sigma U^\dag,~U=\bigoplus_{i=1}^{N_A}\frac{1}{\sqrt{2}}\mqty(1&\im\\1&-\im). 
    \label{eq:U}
\end{align}
Observing that $q_i$ and $p_i$ correspond to $\tilde{S}_i^x$ and $\tilde{S}_i^y$ via the Holstein-Primakoff transformation~\eqref{SMeq:HP}, we obtain $\boldsymbol{s}_i=\boldsymbol{0}$ since we have chosen $(\theta_t,\phi_t)$ such that $\langle \tilde{S}_i^x\rangle=\langle \tilde{S}_i^y\rangle =0$. 
On the other hand, the correlators in $\Sigma$ involving these operators may have non-zero expectation values. 
In particular, the translational symmetry of the system allows us to rewrite the entries of the covariance matrix $\Sigma$ in the Fourier representation, 
\begin{align}
    \Sigma_{ij}
    =
    \frac{1}{N}\sum_\mathbf{k}e^{-\im \mathbf{k}\cdot(\mathbf{r}_i-\mathbf{r}_j)}g_\mathbf{k}, 
    \label{eq:CM}
\end{align}
where $g_\mathbf{k}$ is the $2\times 2$ matrix whose entries are given by 
\begin{align}
   g_{{\bf k}}=\left(\begin{array}{cc} g_{\bf k}^{qq} & g_{\bf k}^{qp}\\g_{\bf k}^{pq} & g_{\bf k}^{pp}\end{array}\right),\quad 
    g_\mathbf{k}^{xy}=\langle \tilde{x}_\mathbf{k} \tilde{y}_\mathbf{-k}+\tilde{y}_\mathbf{k} \tilde{x}_\mathbf{-k}\rangle,~x,y\in\{q,p\}, 
\end{align}
and $\tilde{q}_{\bf k}$ an $\tilde{p}_{\bf k}$ are the operators 
\begin{align}
    \tilde{q}_\mathbf{k}=\frac{b_\mathbf{k}+b_\mathbf{-k}^\dag}{\sqrt{2}},~
    \tilde{p}_\mathbf{k}=\frac{b_\mathbf{k}-b_\mathbf{-k}^\dag}{\im\sqrt{2}}. 
\end{align}
According to Eq.~\eqref{eq:CM}, all the information about $\rho_A$ is encoded in the symbol $g_\mathbf{k}$. 
From the Heisenberg equations of motion $\im \partial_t b_\mathbf{k}=[b_\mathbf{k},\tilde{H}]$, we derive the equations describing the time evolution of $g_\mathbf{k}$,
\begin{align}
    s\partial_t g_\mathbf{k}^{qq}
    &=2(\xi_\mathbf{k}-\kappa_\mathbf{k})g_\mathbf{k}^{qp},
    \\
    s\partial_t g_\mathbf{k}^{qp}
    &=
    -(\xi_\mathbf{k}+\kappa_\mathbf{k})g_\mathbf{k}^{qq}
    +
    (\xi_\mathbf{k}-\kappa_\mathbf{k})g_\mathbf{k}^{pp},
    \\
    s\partial_t g_\mathbf{k}^{pp}
    &=
    -2(\xi_\mathbf{k}+\kappa_\mathbf{k})g_\mathbf{k}^{qp}.
\end{align}
In the initial tilted ferromagnetic state, we have $g_\mathbf{k}^{xy}(t=0)=\delta_{x,y}$. The solution of the equations with these initial conditions is 
\begin{align}
    g_\mathbf{k}^{qq}
    &=
    1-\frac{2\kappa_\mathbf{k}(\xi_\mathbf{k}-\kappa_\mathbf{k})}{\omega_\mathbf{k}^2}\sin^2\qty(\frac{\omega_\mathbf{k}t}{s}),
    \label{eq:qq}
    \\
    g_\mathbf{k}^{qp}
    &=-\frac{2\kappa_\mathbf{k}}{\omega_\mathbf{k}}
    \sin(\frac{\omega_\mathbf{k}t}{s})
    \cos(\frac{\omega_\mathbf{k}t}{s}),
    \label{eq:qp}
    \\
    g_\mathbf{k}^{pp}
    &=
    1+\frac{2\kappa_\mathbf{k}(\xi_\mathbf{k}+\kappa_\mathbf{k})}{\omega_\mathbf{k}^2}
    \sin^2\qty(\frac{\omega_\mathbf{k}t}{s}), 
    \label{eq:pp}
\end{align}
where $\omega_\mathbf{k}=\sqrt{\xi_\mathbf{k}^2-\kappa_\mathbf{k}^2}$.
The each entry of $\Sigma$ and, accordingly, that of $\Gamma$, is calculated by inserting Eqs.~\eqref{eq:qq}, \eqref{eq:qp}, and \eqref{eq:pp} into Eq.~\eqref{eq:CM}. 
We can also obtain from the previous expressions the mode occupation number of spin waves, reported in Eq.~(8) of the main text, 
\begin{align}
    n_\mathbf{k}(t)=\frac{g_\mathbf{k}^{qq}+g_\mathbf{k}^{pp}-2}{4}
    =
    \frac{\kappa_\mathbf{k}^2}{\omega_\mathbf{k}^2}\sin^2\qty(\frac{\omega_\mathbf{k}t}{s}).
    \label{eq:n_k_sm}
\end{align}

\section{Validity of the time-dependent spin-wave theory}\label{sec:validity}
\begin{figure}
    \centering
    \includegraphics[width=0.5\linewidth]{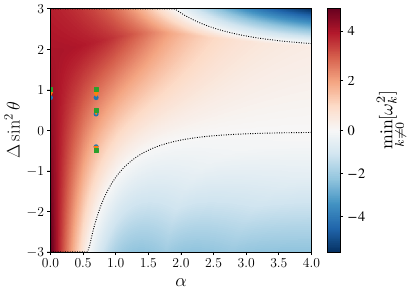}
    \caption{Color map of $\min_{k\neq0}[\omega_k^2]$ in Eq.~\eqref{eq:w_k} as a function of $\Delta \sin^2\theta$ and $\alpha$. The dotted curves are the boundary between the red and blue regions where $\Im[\omega_k]=0$ for all $k$ and $\Im[\omega_k]\neq0$ for certain $k$, respectively. The symbols correspond to the specific parameters used in Fig.~2 of the main text. We take system size $N=24$.}
    \label{fig:DI}
\end{figure}

As mentioned in the main text, the spin-wave theory is valid only when $\Im[\omega_\mathbf{k}]=0$ for all $\mathbf{k}$. Indeed, if there is a mode $\mathbf{k}$ such that $\Im[\omega_\mathbf{k}]\neq0$, then its occupation increases exponentially  in time, as shown in Eq.~\eqref{eq:n_k}. 
To discuss more quantitatively when this condition is met, let us focus on the one-dimensional long-range spin-1/2 XXZ Hamiltonian with power-law decaying couplings, 
\begin{align}
    H_{{\rm XXZ}}
    =
    -\sum_{i\neq j}
    \frac{\hat{\sigma}_i^x \hat{\sigma}_j^x+\hat{\sigma}_i^y \hat{\sigma}_j^y+(1-\Delta)\hat{\sigma}_i^z \hat{\sigma}_j^z}{\abs{\mathbf{r}_i-\mathbf{r}_j}^\alpha K}, 
    \label{eq:XXZ_chain_sp}
\end{align}
where $\hat{\sigma}_i^\mu$ are the standard Pauli matrices on the $i$-th site and $K=N^{-1}\sum_{i\neq j}\abs{\mathbf{r}_i-\mathbf{r}_j}^{-\alpha}$ is the Kac rescaling factor. This is the model used in the main text to 
numerically check our analytic results.
For the Hamiltonian~\eqref{eq:XXZ_chain_sp}, $\omega_\mathbf{k}$ reduces to 
\begin{align}
    \omega_k=2\sqrt{(1-J_k(\alpha))[1-(1-\Delta \sin^2 \theta )J_k(\alpha)]}, \label{eq:w_k}
\end{align}
where $\theta$ is the polar angle of the initial tilted-ferromagnetic state and 
\begin{align}
J_k(\alpha)=\frac{\sum_{j=1}^{N-1} e^{-\im kj}\min(j,N-j)^{-\alpha}}{\sum_{j=1}^{N-1} \min(j,N-j)^{-\alpha}}.
\end{align}
In Fig.~\ref{fig:DI}, we show a color map of $\min_{k\neq 0}[\omega_{k}^2]$ as a function of $\Delta \sin^2 \theta$ and $\alpha$. 
In this figure, the dotted curves delimit the red region where the condition $\Im[\omega_k]=0$ for all $k$ is satisfied and the spin wave theory is valid. By contrast, in the blue regions $\Im[\omega_k]\neq0$ for certain $k$ and the spin-wave theory is not applicable.
As expected, the region of the parameter space where $\mathrm{Im}[\omega_k] = 0$ for all $k$ (red region) expands as $\alpha$ decreases, since the precessing ferromagnetic order is increasingly stabilized as the spin-spin interactions become more long ranged. 
We also find that, for $\Delta \geq -1$, it is sufficient to take $\alpha \geq 1$ for the spin-wave theory to be valid. 
The parameters used in Fig.~2 of the main text correspond to the symbols in Fig.~\ref{fig:DI}, confirming that they lie within the region where $\Im[\omega_k]=0$ for all $k$. 

\section{Derivation of Eq.~(10)}\label{SMsec:asymmetry}

Here, we derive the analytic expressions for the entanglement asymmetry given in Eq.~(10) of the main text by applying bosonic Gaussian state techniques.
To exploit the Gaussianity of $\rho_A$ in calculating the entanglement asymmetry, we first rewrite Eq.~(3) of the main text using the Fourier representation of the projection operator, 
\begin{equation}
\Pi_m=\int_{-\pi}^\pi \frac{\dd \alpha}{2\pi}e^{\im \alpha (\hat{M}_A^z-m)},
\end{equation}
as 
\begin{align}
    \Delta S_A = -\log(\int_{-\pi}^\pi \frac{\dd \alpha}{2\pi} \frac{Z(\alpha)}{Z(0)}), 
    \label{eq:EA_CM}
\end{align}
where $Z(\alpha)$ is the charged moment, defined as 
\begin{align}\label{SMeq:z_alpha}
    Z(\alpha)=\Tr_A\qty[\rho_Ae^{\im \alpha \hat{M}_A^z} \rho_A e^{-\im \alpha \hat{M}_A^z}].
\end{align}
If we rewrite the subsystem magnetization $\hat{M}_A^z$ in the rotated frame, it reads 
\begin{align}
    \hat{M}_A^z=\sum_{i\in A}(\tilde{S}_i^z \cos \theta - \tilde{S}_i^x\sin\theta ). 
    \label{eq:Q_A_rotated_frame}
\end{align}
Applying the Holstein-Primakoff transformation \eqref{SMeq:HP} in Eq.~\eqref{eq:Q_A_rotated_frame}, we obtain
\begin{align}
    \hat{M}_A^z
    &\simeq -s\cos\theta\sum_{i\in A} 
    \qty[ \qty(\frac{q_i}{\sqrt{2s}}+\frac{\tan \theta}{\sqrt{2}})^2+\frac{p_i^2}{2s}-1-\frac{\tan^2\theta}{2}-\frac{1}{2s}].
\end{align}
Introducing the displacement operator 
\begin{align}
    D_A=\exp(-\im \sqrt{\frac{s}{2}}\tan \theta\sum_{i\in A} p_i),
    \label{eq:D}
\end{align}
it diagonalizes $\hat{M}_A^z$ as 
\begin{gather}\label{SMeq:diag_QA}
    \hat{M}_A^z=D_A^\dag \tilde{M}_A^z D_A+N_A\left[\left(s+\frac{1}{2}\right)\cos\theta+s\frac{\sin^2\theta}{2\cos\theta}\right]I,
\end{gather}
where 
\begin{align}
    \tilde{M}_A^z=-\frac{\cos \theta}{2}\sum_{i\in A}(q_i^2 + p_i^2).
    \label{eq:Q_A diagonal}
\end{align}

Applying the result in Eq.~\eqref{SMeq:diag_QA} in Eq.~\eqref{SMeq:z_alpha}, the charged moment can be written as 
\begin{align}
    Z(\alpha)\simeq 
    \Tr_A[\rho_{A,\alpha}\rho_{A,-\alpha}]. 
\end{align}
where $\rho_{A,\pm \alpha}$ stand for 
\begin{align}
    \rho_{A,\pm\alpha}
    =
    e^{\pm \im \frac{\alpha \cos \theta}{2}\hat{n}_A}D_A 
    \rho_A D_A^\dag 
    e^{\mp \im \frac{\alpha \cos \theta}{2}\hat{n}_A}, 
    \label{eq:ChargedMoment}
\end{align}
and
\begin{align}
    \hat{n}_A=\sum_{i\in A} b_i^\dag b_i. 
\end{align}
Note that $\rho_{A,\pm \alpha}$ are also Gaussian states because $D_A$ is a Gaussian operator and $\hat{n}_A$ is quadratic in $\boldsymbol{b}_i$. 
Therefore, Eq.~\eqref{eq:ChargedMoment} shows that, in the time-dependent spin wave approximation, the charged moment $Z(\alpha)$ corresponds to the overlap between two bosonic Gaussian states. 
As explicitly shown in Appendix B of Ref.~\cite{yamashika-2025}, this property allows us to derive, applying the Wigner function formalism, an explicit expression of the charged moment \eqref{eq:ChargedMoment} in terms of the mean vector $\boldsymbol{s}_{\pm \alpha}$ and the covariance matrix $\Sigma_{\pm \alpha}$ of $\rho_{A, \pm\alpha}$, 
\begin{align}
    \frac{Z(\alpha)}{Z(0)}
    =
    e^{-N_A F(\alpha)}
    \label{eq:Z/Z}
\end{align}
where 
\begin{align}
    N_AF(\alpha)
    =
    (\boldsymbol{s}_\alpha-\boldsymbol{s}_{-\alpha})^T
    (\Sigma_\alpha+\Sigma_{-\alpha})^{-1}
    (\boldsymbol{s}_\alpha-\boldsymbol{s}_{-\alpha})
    +
    \frac{1}{2}\log(\det[\frac{\Sigma_\alpha+\Sigma_{-\alpha}}{2\Sigma}]).
\end{align}
The entries of $\boldsymbol{s}_{\alpha}$ and $\Sigma_{\alpha}$ are 
\begin{align}
    [\boldsymbol{s}_{\alpha}]_i=\sqrt{s}\tan\theta 
    \mqty(\cos(\alpha \cos \theta/2)\\ \sin(\alpha \cos \theta/2)),\quad 
    [\Sigma_{\alpha}]_{ij}
    =
    e^{-\im \sigma_y\alpha \cos \theta/2 }
    \Sigma_{ij}
    e^{\im \sigma_y \alpha \cos \theta/2 }. 
\end{align}
\par 
Inserting Eq.~\eqref{eq:Z/Z} in Eq.~\eqref{eq:EA_CM}, we obtain 
\begin{align}
    \Delta S_A = -\log(\int_{-\pi}^\pi \frac{\dd \alpha}{2\pi}e^{-N_A F(\alpha)}). 
    \label{eq:EA_exp}
\end{align}
For $N_A\gg1$, the integral in $\alpha$ can be evaluated using the saddle point approximation. 
The saddle point condition for the integrand, $\partial_\alpha F(\alpha)=0$, yields the saddle point at $\alpha=0$ and, therefore, the integral in Eq.~\eqref{eq:EA_exp} can be approximated as 
\begin{align}
    \Delta S_A \simeq 
    -\log(\int_{-\infty}^{\infty} \frac{\dd \alpha}{2\pi}e^{-\frac{N_A\alpha^2}{2}[\partial_\alpha^2 F(\alpha)]_{\alpha=0}}), 
    \label{eq:EA_Gaussian}
\end{align}
where $[\partial_\alpha^2 F(\alpha)]_{\alpha=0}$ in the leading order in $N_A$ reads 
\begin{align}
    [\partial_\alpha^2 F(\alpha)]_{\alpha=0}\simeq 
    \frac{s\sin^2\theta (\boldsymbol{v}^T \Sigma^{-1}\boldsymbol{v})}{N_A},  
\end{align}
with $\boldsymbol{v}=(0,1,0,1,...,0,1)^T$. 
Within this approximation, the integral in Eq.~\eqref{eq:EA_Gaussian} can be calculated by the standard Gaussian integration techniques, resulting in 
\begin{align}
    \Delta S_A 
    \simeq 
    \frac{1}{2}
    \ln(2\pi s \sin^2 \theta [\boldsymbol{v}^T \Sigma^{-1}\boldsymbol{v}]).
    \label{eq:EA_Sigma}
\end{align}
Recalling $\Gamma=U\Sigma U^\dag$, where $U$ is given in Eq.~\eqref{eq:U}, one finds that Eq.~\eqref{eq:EA_Sigma} is equal to Eq.~(10) of the main text.

\section{Derivation of Eq.~(11)}\label{SMeq:expansion_asymm}

Here, we derive Eq.~(11) of the main text from Eq.~\eqref{eq:EA_Sigma}, identifying the explicit form of the sub-leading term of order $O(n_{\mathbf{k} \neq 0}/n_{\mathbf{0}})$.
\par 
Let us first decompose the covariance matrix $\Sigma$ into the contributions of $\mathbf{k=0}$ and $\mathbf{k\neq0}$ modes as 
\begin{align}
    \Sigma=\Sigma^{(\mathbf{0})}+\Sigma^{(\mathbf{k\neq0})}, 
\end{align}
where the entries of $\Sigma^{(\mathbf{0})}$ and $\Sigma^{(\mathbf{k\neq0})}$ are given by 
\begin{align}
    \Sigma^{(\mathbf{0})}_{ij}
    =
    \mqty(
    \delta_{i,j}
    & 
    -\frac{2\kappa_\mathbf{0} t}{sN}
    \\
    -\frac{2\kappa_\mathbf{0} t}{sN}
    & 
    \delta_{i,j}+N\qty(\frac{2\kappa_\mathbf{0} t}{sN})^2 
    ),\quad 
    \Sigma^{(\mathbf{k\neq0})}_{ij}
    =
    \frac{1}{N}
    \sum_\mathbf{k\neq0}
    e^{-\im \mathbf{k}\cdot(\mathbf{r}_i-\mathbf{r}_j)}
    (g_\mathbf{k}-I).
\end{align}
Since the mode $\mathbf{k=0}$ is dominant except at short times as shown in Eq.~(13), we expand $\boldsymbol{v}^T \Sigma^{-1} \boldsymbol{v}$ in terms of $\Sigma^{(\mathbf{k\neq0})}$ as follows 
\begin{align}
    \boldsymbol{v}^T \Sigma^{-1} \boldsymbol{v}
    &=
    \boldsymbol{v}^T (I+[\Sigma^{(\mathbf{0})}]^{-1} \Sigma^{(\mathbf{k\neq 0})})^{-1}(\Sigma^{(\mathbf{0})})^{-1} \boldsymbol{v}
    \nonumber\\
    &= 
    \boldsymbol{v}^T (\Sigma^{(\mathbf{0})})^{-1} \boldsymbol{v}
    -
    \boldsymbol{v}^T (\Sigma^{(\mathbf{0})})^{-1}
    \Sigma^{(\mathbf{k\neq 0})}
    (\Sigma^{(\mathbf{0})})^{-1}
    \boldsymbol{v}
    +O[(\Sigma^{(\mathbf{k\neq 0})})^2].
    \label{eq:vSv expand}
\end{align}
By simple algebra, one finds that the inverse of $\Sigma^{(\mathbf{0})}$ is
\begin{align}
    [\Sigma^{(\mathbf{0})}]^{-1}_{ij}
    =
    \mqty(
    \delta_{i,j}
    +
    \frac{4n_\mathbf{0}f_A}{N[1+4n_\mathbf{0}f_A(1-f_A)]}
    &
    \frac{2\kappa_\mathbf{0}t}{sN[1+4n_\mathbf{0}f_A(1-f_A)]}
    \\
    \frac{2\kappa_\mathbf{0}t}{sN[1+4n_\mathbf{0}f_A(1-f_A)]}
    &
    \delta_{i,j}
    -
    \frac{4n_\mathbf{0}f_A}{N[1+4n_\mathbf{0}f_A(1-f_A)]}
    ), 
    \label{eq:Sigma inverse}
\end{align}
where $n_\mathbf{0}=\kappa_\mathbf{0}^2(t/s)^2$ is the number of spin waves with zero momentum. 
Inserting Eq.~\eqref{eq:Sigma inverse} into Eq.~\eqref{eq:vSv expand}, we obtain 
\begin{align}
    \boldsymbol{v}^T \Sigma^{-1}\boldsymbol{v}
    \simeq 
    \frac{N_A(1+X_A)}{1+4n_\mathbf{0}f_A(1-f_A)}. 
    \label{eq:vSv_2}
\end{align}
Here, $X_A$ is the contribution of the $\mathbf{k\neq0}$ modes and is specifically given by 
\begin{multline}
    X_A
    =
    \frac{f_A}{1+4n_\mathbf{0}f_A(1-f_A)}
    \sum_\mathbf{k\neq0}\abs{\delta^A_{\mathbf{k},\mathbf{0}}}
    \Bigg[
    \frac{4n_\mathbf{0}f_A^2\kappa_\mathbf{k}(\xi_\mathbf{k}-\kappa_\mathbf{k})}{\omega_\mathbf{k}^2}
    \sin^2\qty(\frac{\omega_\mathbf{k}t}{s})
    \\
    +
    \frac{2\kappa_\mathbf{0}\kappa_\mathbf{k}f_A (t/s)}{\omega_\mathbf{k}}
    \sin(\frac{\omega_\mathbf{k}t}{s})
    \cos(\frac{\omega_\mathbf{k}t}{s})
    -
    \frac{\kappa_\mathbf{k}(\xi_\mathbf{k}+\kappa_\mathbf{k})}{\omega_\mathbf{k}^2}\sin^2\qty(\frac{\omega_\mathbf{k}t}{s})
    \Bigg],
    \label{eq:X_A}
\end{multline}
with $\delta_{\mathbf{k,0}}^A=N_A^{-1}\sum_{i\in A} e^{\im\mathbf{k}\cdot \mathbf{r}_i}$. 
Applying Eq.~\eqref{eq:vSv_2} in Eq.~\eqref{eq:EA_Sigma} and expanding the latter in terms of $X_A$, we find
\begin{align}
    \Delta S_A 
    \simeq 
    \frac{1}{2}\ln(\frac{2s N_A\sin^2\theta}{1+4n_\mathbf{0}f_A(1-f_A)})
    +X_A.
    \label{SMeq:EA_analytic_full}
\end{align}
\par
\begin{figure}[t]
    \raggedright
    \includegraphics[width=1\linewidth]{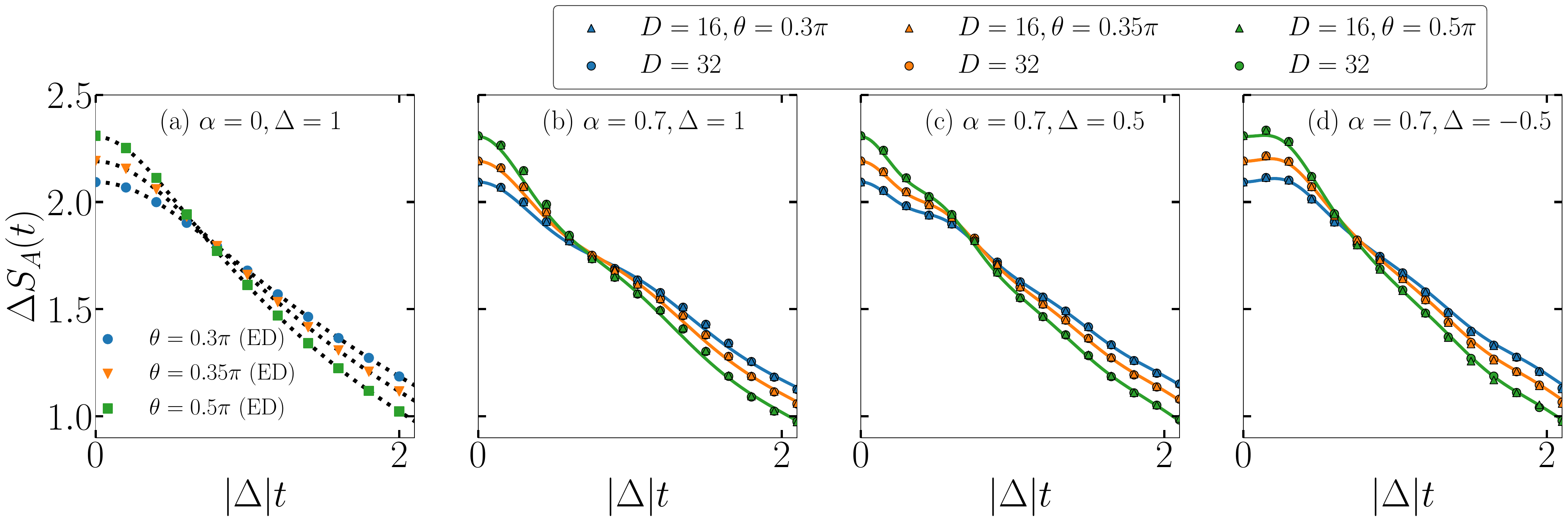}
    \caption{Time evolution of the entanglement asymmetry from different initial tilted ferromagnetic states in the long-range spin-$1/2$ chain~\eqref{eq:XXZ_chain_sp}. (a) The symbols are the exact asymmetry calculated with ED using Eq.~\eqref{eq:EA_exact} and the dashed curves correspond to Eq.~\eqref{SMeq:EA_analytic_full}. (b)-(d) The symbols were obtained with the MPS-TDVP algorithm for maximal bond-dimension $D=16$ (triangles) and $D=32$ (circles). The solid lines are the analytic prediction~\eqref{SMeq:EA_analytic_full} of the spin-wave theory taking into account the contributions of the $\mathbf{k\neq0}$ modes, contained in the $X_A$ term. We set $N=128$ and $N_A=32$ in all the plots.}
    \label{fig:EA_SM}
\end{figure}

 
This is Eq.~(11) of the main text. Note that $X_A$ corresponds to the unspecified $O(n_\mathbf{k\neq0}/n_\mathbf{0})$ term in that equation, which can be explicitly computed using Eq.~\eqref{eq:X_A}. In Fig.~2 of the main text, we compared Eq.~\eqref{SMeq:EA_analytic_full}, excluding the term $X_A$, withexact numerical results obtained with ED. We attributed the deviations at short times to the neglect of the contribution $X_A$. In Fig.~\ref{fig:EA_SM}, we now compare the analytic prediction~\eqref{SMeq:EA_analytic_full} including the term $X_A$ with the numerical results obtained using the matrix-product-state time-dependent-variational-principle (MPS-TDVP) (for two different elections of the bond dimension, see Sec.~\ref{SMsec:numerics} for details), showing a good agreement between them even at short times. 

\section{ Numerical results for large systems}\label{SMsec:numerics}    
 
While the numerical results presented in the main text correspond to experimentally accessible small systems, for completeness we provide here numerical results for larger system sizes that are out of reach of brute-force ED. As in the main text, we consider the one-dimensional long-range spin-$1/2$ XXZ Hamiltonian~\eqref{eq:XXZ_chain_sp}. 
In the case $\alpha=0$, one can exploit the specific form of the Hamiltonian to optimize ED and reach larger system sizes. For $\alpha>0$, we apply the matrix-product-state time-dependent-variational-principle (MPS-TDVP) method.

\subsection{Exact Diagonalization for \texorpdfstring{$\alpha=0$}{alpha=0}}

In the long-range interaction limit $\alpha\to0$, the Hamiltonian \eqref{eq:XXZ_chain_sp} can be written in terms of total magnetization operators as 
\begin{align}
    H_{{\rm XXZ}}
    =
    \frac{4\Delta (\hat{M}^z)^2}{N}
    -\frac{4\hat{\mathbf{M}}^2}{N}.
    \label{eq:XXZ_a=0}
\end{align}
Since the initial-tilted ferromagnetic state~(1) is an eigenstate of $\hat{\mathbf{M}}^2$ and $[H_{{\rm XXZ}},\hat{\mathbf{M}}^2]=0$, $\hat{\mathbf{M}}^2$ is conserved during the dynamics. 
We thus neglect $\hat{\mathbf{M}}^2$ in Eq.~\eqref{eq:XXZ_a=0} in the following. 
\par 
It is convenient to introduce the computational basis $\ket{\mathbf{n}}=\bigotimes_{i=1}^N \ket{n_i}$ ($n_{i}=0,1$) such that $\hat{\sigma}_i^z \ket{\mathbf{n}}=(-1)^{n_i}\ket{\mathbf{n}}$.  
In that basis, the tilted-ferromagnetic state can be expanded as 
\begin{align}
    \ket{\Psi_0}
    &=
    \sum_{m=0}^{N}
    \cos^m(\theta/2)
    \sin^{N-m}(\theta/2)
    \sum_{|\mathbf{n}|=N-m}\ket{\mathbf{n}}
    \\
    &=
    \sum_{m=0}^{N}
    \sqrt{\mqty(N\\m)}
    \cos^m(\theta/2)
    \sin^{N-m}(\theta/2)
    \ket{m-N/2}, 
    \label{eq:tilted-ferro_expand}
\end{align}
where 
\begin{align}
    \ket{m-N/2}=\mqty(N\\m)^{-1/2}\sum_{|\mathbf{n}|=N-m} \ket{\mathbf{n}}
\end{align}
is the simultaneous eigenstate of $\hat{M}^z$ and $\hat{\mathbf{M}}^2$ with eigenvalues $m-N/2$ and $N/2(N/2+1)$, respectively. 
Multiplying the time-evolution operator $e^{-\im \frac{\Delta t}{N}(\hat{M}^z)^2 }$ to Eq.~\eqref{eq:tilted-ferro_expand}, we obtain the explicit form of the time-evolved state 
\begin{align}
    \ket{\Psi_t}
    =
    \sum_{m=0}^{N}
    \sqrt{\mqty(N\\m)}
    e^{-\im \frac{4\Delta t}{N}\qty(m-\frac{N}{2})^2}
    \cos^m(\theta/2)
    \sin^{N-m}(\theta/2)
    \ket{m-N/2}. 
    \label{eq:|Psi_t>}
\end{align}
To calculate the reduced density matrix $\rho_A$, we need to perform on Eq.~\eqref{eq:|Psi_t>} the partial trace over the Hilbert space for the complement of $A$. 
To this end, we decompose $\ket{n-N/2}$ in Eq.~\eqref{eq:|Psi_t>} as 
\begin{align}
    \ket{n-N/2}=
    \sum_{n_A=0}^{N_A}
    \sqrt{\frac{\mqty(N_A\\n_A)\mqty(N_{\bar A}\\n_{\bar{A}})}{\mqty(N\\n)}}
    \ket{n_A-N_A/2}\otimes\ket{n_{\bar{A}}-N_{\bar{A}}/2}, 
    \label{eq:|M^z>}
\end{align}
where $N_{\bar{A}}=N-N_A$, $n_A=n-n_{\bar{A}}$, and $\ket{n_{A (\bar{A})}-N_{A(\bar{A})}/2}$ is the eigenstate of $\hat{M}_{A(\bar{A})}^z =\sum_{i\in A(\bar{A})} \hat{S}_i^z$ with eigenvalues $n_{A(\bar{A})}-N_{A(\bar{A})}/2$.
Inserting Eq.~\eqref{eq:|M^z>} into Eq.~\eqref{eq:|Psi_t>} and taking the partial trace, we obtain 
\begin{align}
    \rho_A (t)
    &= 
    \sum_{n_{\bar{A}}=0}^{N_{\bar{A}}} 
    \bra{n_{\bar{A}}-N_{\bar{A}}}\ket{\Psi_t}
    \bra{\Psi_t}\ket{n_{\bar{A}}-N_{\bar{A}}},
    \\
    &=
    \sum_{n_A,n_A'=0}^{N_A}
    C_{n_A,n_A'}(t,\theta)
    \ket{n_A-N_A/2}\bra{n_A'-N_A/2}, 
    \label{eq:rho_A explicit}
\end{align}
where 
\begin{multline}
    C_{n_A,n_A'}(t,\theta)
    =
    \sum_{n,n'=0}^N
    \delta_{n-n_A,n'-n_A'}
    \cos^{n+n'}(\theta/2) \sin^{N-n-n'}(\theta/2)
    \\
    \times
    e^{-\frac{4\im \Delta t}{N}[(n-N/2)^2-(n'-N/2)^2]}
    \sqrt{
    \mqty(N_A\\n_A)
    \mqty(N_A\\ n_A')
    \mqty(N-N_A\\n-n_A)
    \mqty(N-N_A\\n'-n_A')}. 
\end{multline}
This expression allows us to easily calculate the symmetrized reduced density matrix, $\tilde{\rho}_A=\sum_m \Pi_m \rho_A \Pi_m$ where $\Pi_m$ is the projection onto the eigenspace of $\hat{M}^z_A$ with eigenvalue $m$. We obtain
\begin{align}
    \tilde{\rho}_A(t)=\sum_{n_A=0}^{N_A} C_{n_A,n_A}(t,\theta) 
    \ket{n_A-N_A/2}\bra{n_A-N_A/2}. 
    \label{eq:til rho_A}
\end{align}
Plugging Eqs.~\eqref{eq:rho_A explicit} and \eqref{eq:til rho_A} into Eq.~(3) of the main text, we finally arrive at 
\begin{align}
    \Delta S_A(t) 
    =
    \ln( \frac{\sum_{n_A,n_A'=0}^{N_A}C_{n_A,n_A'}(t,\theta) C_{n_A',n_A}(t,\theta)}{\sum_{n_A=0}^{N_A} C_{n_A,n_A}(t,\theta)^2}). 
    \label{eq:EA_exact}
\end{align}
This formula is exact at any $t$ and allows us to compute the entanglement asymmetry for system sizes $N$ far beyond the reach of brute force ED.  
The symbols in Fig.~\ref{fig:EA_SM}~(a) are obtained by computing numerically Eq.~\eqref{eq:EA_exact} with $N=128$ and $N_A=32$. 

\subsection{Matrix-Product-State Time-Dependent-Variational-Principle for $\alpha>0$}

For spatially decaying interactions ($\alpha > 0$), we use the MPS-TDVP algorithm \cite{Jutho-2016, Jutho-Cirac-2011}, which yields quasi-exact time evolution for long-range systems. 
To handle algebraically decaying interactions, we approximate the corresponding matrix-product operator (MPO) as a sum of MPOs for exponentially decaying interactions, adapting a technique from Ref.~\cite{Pirvu_2010}. 
This approximation leverages the periodicity of the interaction strength, $f(l)=|\mathbf{r}_i-\mathbf{r}_{i+l}|^{-\alpha}/K=f(l+N)$, allowing its representation as a sum of exponentials with imaginary exponents. We set the relative tolerance for this approximation as $10^{-10}$, which requires a MPO bond dimension of order $O(N)$.
\par 
Setting a threshold $D$ for the MPS bond dimension, we use the hybrid MPS-TDVP scheme depending on whether the dimension of each bond is larger than $D$ or not: For bonds with dimension below $D$, we use the two-site MPS-TDVP scheme with randomized singular-value decomposition~\cite{mcculloch2024}, truncating Schmidt coefficients smaller than $10^{-10}$, while for bonds with dimension larger than $D$, we employ the standard single-site MPS-TDVP~\cite{Jutho-2016}.
We subsequently increase $D$ until the entanglement asymmetry converges within the desired time window (see Fig.~\ref{fig:EA_SM}). 
The convergence is already achieved for $D=32$.

\end{document}